\documentclass[conference, a4paper,10pt]{IEEEtran}
\IEEEoverridecommandlockouts
\usepackage{times}
\usepackage{cite}
\usepackage{textcomp}
\usepackage[dvips]{color}
\usepackage{epsf}
\usepackage{epsfig}
\usepackage{graphicx}
\usepackage{graphics}
\usepackage{epstopdf}
\usepackage[justification=centering,small]{caption}
\usepackage[bottom]{footmisc}
\usepackage{amsmath}
\usepackage{amssymb}
\usepackage{amsthm}
\usepackage{amsxtra}
\usepackage{algorithm}
\usepackage{algorithmic} 	
\usepackage{enumerate}
\usepackage{multirow}
\usepackage{enumerate}
\usepackage{amssymb}
\usepackage{lipsum}
\usepackage{setspace}
\usepackage{multirow}
\usepackage{tikz}
\usepackage{setspace}
\usepackage{subfigure}

\newtheorem{theorem}{\bf Theorem}

\newtheorem{property}{\textit{Property}}

\usepackage{rotating} 
\usepackage{graphicx} 

\newcommand{\Rmnum}[1]{\expandafter\@slowromancap\romannumeral #1@}

\makeatother

\begin{document}

\title{Synthetic Generation of Solar States for Smart Grid: A Multiple Segment Markov Chain Approach}
\author{\IEEEauthorblockN{Wayes Tushar\IEEEauthorrefmark{1}, Shisheng Huang\IEEEauthorrefmark{1},
Chau Yuen\IEEEauthorrefmark{1}, Jian (Andrew) Zhang\IEEEauthorrefmark{2}, and  David B. Smith\IEEEauthorrefmark{4}}
\IEEEauthorblockA{\IEEEauthorrefmark{1}Singapore University of Technology and Design, Singapore 138682.\\
\IEEEauthorrefmark{2}CSIRO Computational Informatics, Marsfield, NSW, Australia.\\
\IEEEauthorrefmark{4}National ICT Australia (NICTA), Canberra 2601, Australia.\\Email:
\{wayes\_tushar,yuenchau,shisheng\_huang\}@sutd.edu.sg; andrew.zhang@csiro.au; david.smith@nicta.com.au.}
\thanks{This work is supported by the Singapore University of Technology and Design (SUTD) under the Energy Innovation Research Program (EIRP) Singapore NRF2012EWT-EIRP002-045.}
\thanks{\IEEEauthorrefmark{4}David Smith is also with the Australian National University (ANU), and his work is supported by NICTA. NICTA is funded by the Australian Government through the Department of Communications and the Australian Research Council through the ICT Centre of Excellence Program.}
}
\maketitle
\begin{abstract}
The use of photovoltaic (PV) sources is becoming very popular in smart grid for their ecological benefits, with higher scalability and utilization for local generation and delivery. PV can also potentially avoid the energy losses that are normally associated with long-range grid distribution. The increased penetration of solar panels, however, has introduced a need for solar energy models that are capable of producing realistic synthetic data with small error margins. Such models, for instance, can be used to design the appropriate size of energy storage devices or to determine the maximum charging rate of a PV-powered electric vehicle (EV) charging station. In this regard, this paper proposes a stochastic model for solar generation using a Markov chain approach. Based on real data, it is first shown that the solar states are inter-dependent, and thus suitable for modeling using a Markov model. Then, the probabilities of transition between states are shown to be heterogeneous over different time segments. A model is proposed that captures the inter-temporal dependency of solar irradiance through segmentation of the Markov chain across different times of the day. In the studied model, different state transition matrices are constructed for different time segments, which the proposed algorithm then uses to generate the solar states for different times of the day. Numerical examples are provided to show the effectiveness of the proposed synthetic generator.
\end{abstract}
\begin{keywords}
Multiple segment, Markov chains, photovoltaic, smart grid, real data.
\end{keywords}
\IEEEpeerreviewmaketitle
\section{Introduction}\label{sec:introduction}
Significant concerns of today's energy sector consist of issues such as continuous increase in energy demand, fast depletion of conventional energy resources, and the effect on the environment~\cite{Liu-STSP:2014,Tushar-TIE:2014,Naveed-Energies:2013,Naveed-smartgridcomm:2014,ch28imp,chaibo-TSG:2014,Hassan-ICC:2013,Tushar-Globecom:2014,Liu-ISGT:2013}. Solar energy is a great source of renewable and clean energy that has the capability to effectively solve these problems. For example, most of the renewable energy from solar photovoltaics (PVs) at present is either delivered to the grid or to isolated loads, e.g., islanded micro-grids~\cite{Fang-J-CST:2012}, to help meet demand. In fact, various forms of solar energy such as solar heat, solar PV, solar thermal, and solar fuels offer abundant, clean and environmental friendly energy resources that have the potential to address the most compelling energy problems that are faced by the energy sector. Furthermore, the continuing decrease in cost of PV arrays and the increase in their efficiency means there is an even more promising role for PV generating systems in the near future~\cite{Tushar-TIE:2014}.

Due to the potential benefits of solar energy in future smart grids, problems related to the deployment of solar energy sources have been explored by much literature, as surveyed in \cite{Fang-J-CST:2012}. For instance,  the concept of PV generation, and its feasibility for practical application has been studied in \cite{Burns-EnergyPolicy:2013}, whereas \cite{Eftichios-SolarEnergy:2006} investigates the optimal sizing of stand alone photovoltaic systems using genetic algorithms. Modeling of photovoltaic cells is studied in, e.g., \cite{Chatterjee-TEC:2011}, and its use as a distributed energy source has been explored in, e.g., \cite{Wies-TPS:2005, Phuangpornpitak-RSER:2005}, in terms of both experiment and simulation. In \cite{Shisheng-JEE:2012}, the authors explore the utility level benefits of distributed PV systems that are coupled with electricity storage. The possibility of using a solar power system in combination with other different forms of generation, e.g., wind turbines, is studied in  \cite{Nabil-ECM:2008}. Finally, studies such as \cite{Kottas-TEC:2006} have focussed on designing schemes/protocols that maximize the power output of PV generators.

Apart from the above, another important issue is modeling the synthetic generation of renewable energy, which has recently gained considerable attention. Modeling synthetic generation of solar energy is of paramount importance. Depending on the local irradiance states, the instantaneous  power output of a solar generator can vary significantly. Therefore, it raises a concern as to the power quality of the electricity grid as $90\%$ of installed PV systems worldwide are grid connected~\cite{Dong-Energy:2013}. Furthermore, with the advancement of smart grids, solar panels have started to be installed as distributed generators not just for household use, but also for heavy electric loads such as charging of EVs in vehicle charging stations and residential premises~\cite{Yousuf-saber-TIE:2011}. To this end, there is a need for energy consumers to understand the output of their solar irradiance states based on their solar energy usage In fact, devising accurate models that can generate solar irradiance states over time can significantly help electricity users to make decisions on when to use the generated solar energy, and for what type of work they need to use the energy for. For instance, if the intensity of the solar irradiance is very low, depending on the type of solar panels the amount of energy that would be generated might also be very low to accomplish any meaningful work. On the contrary, when the solar irradiance state is very high, the likelihood of very high generation of solar energy will be large, and thus this will enable the users to use the generated energy for heavy loads such as EV charging. In this regard, there is a need for accurate synthetic data generator models that can generate different solar states, e.g., low, medium or high, over time to not only allow the grid operator to implement on-the-fly grid management systems~\cite{Dong-Energy:2013}, but also to help consumers in making energy usage decisions, such as what is the appropriate size of storage device to store energy, or what is the maximum EV charging rate at any particular time of the day.

Due to the fact that the states of solar irradiance can randomly change from low to high or vice-versa during a day, we take the first step in this paper to use a Markov model to design a synthetic solar data generator. We classify the solar states into four categories including low, medium, high and very high depending on their application, as described in the next section, and propose a segmented first order Markov chain model to capture the transition between states. We note that Markov chain models have been used extensively to generate synthetic wind data~\cite{Ettoumi-RenewableEnergy:2003}. However, their use is not well explored for generating synthetic solar data due to the fact, as we will see later, that the transition probability of solar states is not stationary over time. In this paper, we effectively capture this temporal variation by designing different state transition matrices~\cite{Ettoumi-RenewableEnergy:2003}, for different time segments of the day.  Using real solar data, we differentiate between the four solar states by including a separate range of values of solar irradiation in each state. Depending on the range of values in each state, such a general classification would facilitate decision making of users as to when to use their PV energy for different purposes, irrespective of location and time.

To this end, we first show that the solar states are interdependent, and therefore the Markov model is a reasonable model to generate synthetic solar states. Then, we demonstrate that the transition probabilities of solar states are not stationary over times. With a view to capture the time dependency, we divide the entire solar irradiance time duration of interest into multiple segments, and model different state transition matrices for each time segment. We propose an algorithm to generate the synthetic solar state at each time-instant of interest based on the state transition probability. To show the effectiveness of the proposed generator, we compare the result with real measured solar states, and compute the affinity towards the real measured states in terms of their standard deviation. The proposed synthetic solar state generator can be used in any area of interest by simply computing the segmented transition probability matrices for that area's specific time series data set.
\section{Solar Generation as Markov Chains}
A Markov chain represents a stochastic process in which a state changes at discrete time steps. Mathematically, a first order Markov chain is a sequence of random variables $(x_1, x_2,\hdots, x_N)$ such that a future state $x_{n+1}$ depends conditionally upon the current state $x_n$ only, and is independent of all past states $(x_1, x_2,\hdots, x_{n-1})$. A Markov chain is modeled in terms of its transition probabilities $p_{i,j}$, and a first order transition probability matrix $\mathbf{P}$, $p_{i,j}$ determines the probability of transitioning of a state from $i$ to $j$ regardless of previous states that were visited~\cite{Brokish-PSCE:2009}.

We note that the solar irradiance in a particular area is highly dependent on its geographic location and is subject to change from one time to the next based on weather conditions at different times of the day. For example, in a normal sunny day, the solar irradiance is lower in early morning, eventually increases with time to reach its peak at noon, and finally again gradually diminishes as the afternoon progresses. Since the change of solar states from one time instant to the next is gradual rather than abrupt, it is reasonable to model the random generation of solar power in terms of a Markov chain to capture its time correlation between states.

For that purpose, we represent solar data with a first order Markov chain approach. We define a solar irradiance intensity state space $X$ where each state $i\in X$ of the chain refers to a range of solar radiation intensity $I_s$ in Watt per square meter ($W/m^2$). For any given state $i$ there is a probability $p_{i,j}$ of what the next state $j$ will be, as noted above. We stress that, for different case studies, the discretization of irradiance states can be different in accordance with the chosen method and data sets. However, in this paper, we are more interested in observing the solar irradiance from a general point of view, such as high or low irradiance, based on some predefined threshold ranges. In this regard, we consider that the state space of the stochastic process $X(t)$ modeling the solar intensity consists of four discrete states
\begin{eqnarray}
X = \{ x_\text{L}, x_\text{M}, x_\text{H}, x_\text{VH}\},
\label{eqn:states}
\end{eqnarray}
\begin{table}[b!]
\centering
\small
\caption{Interpretation of different solar states in the Markov chain.}
\begin{tabular}{|c|c|c|}
\hline
State & Range of irradiance & Suitable applications \\
\hline\hline
$x_\text{L}$ & $I_{s}\leq I_{s,\text{lr}}$ & Not suitable for any meaningful application \\
\hline
$x_\text{M}$ &  $I_{s,\text{lr}}<I_{s}\leq I_{s,\text{mr}}$ & Light energy consuming loads \\
\hline
$x_\text{H}$ & $I_{s,\text{mr}}<I_{s}\leq I_{s,\text{max}}$ & High household loads \\
\hline
$x_\text{VH}$ &  $I_{s}> I_{s,\text{max}}$ & Very high loads (e.g., EV charging)\\
\hline
\end{tabular}
\label{table:1}
\end{table}
where each state $i\in X$ refers to a range of solar intensities in $W/m^2$. To differentiate between states, we consider three threshold values $I_{s,\text{lr}}, I_{s,\text{mr}}$ and $I_{s,\text{max}}$. We assume that when $I_s<I_{s,\text{lr}}$, the solar irradiance is very low, and hence is not suitable for generating enough solar energy to do any meaningful task. When $I_{s,\text{lr}}<I_{s}\leq I_{s,\text{mr}}$, the generated solar energy from this irradiance has enough strength to perform light household tasks such as charging batteries, running televisions, and operating energy-saving bulbs and fans. The ranges $I_{s,\text{mr}}<I_{s}\leq I_{s,\text{max}}$ and $I_{s}> I_{s,\text{max}}$ refer to  high $x_\text{H}$ and very high $x_\text{VH}$ states respectively. When the solar state is high, it is assumed that users can use the produced energy for high energy consuming household appliances such as washing machines and water heaters. Finally, when the solar irradiance state is at $x_\text{VH}$, the produced solar energy is enough to be used for very high loads such as charging electric vehicles or running air conditioners. We summarize the irradiance range of each defined state, and their suitability for different applications in Table~\ref{table:1}. It is important to note that the number of states in Table~\ref{table:1} can be increased by setting a different ranges of values from the time series data within each state, and that this increases computational complexity. However, to generalize the model irrespective of the location, we assume that $I_{s,\text{lr}}, I_{s,\text{mr}}$ and $I_{s,\text{max}}$ are always fixed\footnote{That is, irrespective of the location e.g., whether in USA or in Australia, the thresholds have the same value.} for the defined states in \eqref{eqn:states}.

\begin{figure}[t!]
\centering
\includegraphics[width=0.8\columnwidth]{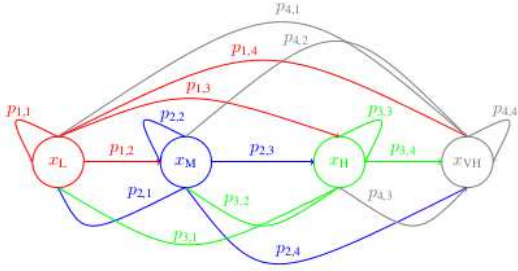}
\caption{Probability of transition between different states of Markov chain. The transition probability $p_{i,j}$ from state $i$ to $j$ for each pair $(i,j)$ is written just above the transition line between the pair of states.} \label{fig:figure-1}
\end{figure}

Now, for a considered sequence of time instants $t_1<t_2<\hdots <T$, where $T$ is the total time of solar measurement and $t$ is the index for each time, the conditional probability that the solar state $i$ at time $t$ will switch to state $j$ at $t+1$ is
\begin{eqnarray}
p_{i,j}\{X(t+1) = {j}|X(t) = {i}\};~ i, {j}\in X.
\end{eqnarray}
To this end, for the considered four states in \eqref{eqn:states}, we define a first order state transition probability matrix of size $4\times 4$ as follows:
\begin{eqnarray}
\mathbf{P}=\begin{bmatrix}
p_{1,1} & p_{1,2} & p_{1,3} & p_{1,4}\\
p_{2,1} & p_{2,2} & p_{2,3}& p_{2,4}\\
p_{3,1} & p_{3,2} & p_{3,3}& p_{3,4}\\
p_{4,1} & p_{4,2} & p_{4,3} & p_{4,4}\\
\end{bmatrix}.
\label{eqn:eqn-1}
\end{eqnarray}
In \eqref{eqn:eqn-1}, each $p_{i,j}\in\mathbf{P}$ is the probability of transition of a state $i$ at time slot $t$ to a state $j$ in the next time slot $t+1$, where each $i,j\in\{1,2,3.4\}$ is assumed to have a one-to-one relationship with states $\{x_\text{L}, x_\text{M}, x_\text{H}, x_\text{VH}\}$. We graphically explain the probabilities of transitions between states in Fig.~\ref{fig:figure-1}.
\subsection{Transition Probability Matrix}
\begin{figure}[t!]
\centering
\includegraphics[width=\columnwidth]{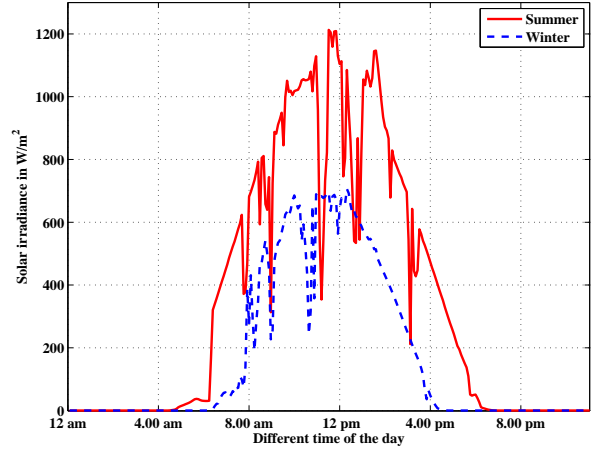}
\caption{Solar irradiance data for a typical day, both in summer and in winter, in Colorado, USA during 2013.} \label{fig:solar-data}
\end{figure}
The analysis of solar irradiance is carried out over a time series data set that contains the values of solar irradiance for every five minutes during one month of summer\footnote{Similar analysis is equally applicable for time series data sets for any other season.} in Colorado, USA in 2013~\cite{NREL-2014}. (An example solar irradiance for a typical day, both in summer and in winter, is shown in Fig.~\ref{fig:solar-data}). Now for analysis, first the set of data is converted into four solar irradiance states according to Table \ref{table:1}, where we consider $I_{s,\text{lr}} = 200$, $I_{s,\text{mr}} = 450$ and $I_{s,\text{max}} = 500$ W/m$^2$. Then, from the state transition matrix,
\begin{eqnarray}
\mathbf{M}_s=\begin{bmatrix}
 52 & 1 & 0 & 0\\
 1 & 38 & 1 & 0\\
0 & 1 & 5 & 1\\
0 & 0 & 1 & 79\\
\end{bmatrix},
\label{eqn:state-transition-matrix-2}
\end{eqnarray}
of the given data set, the first order $4\times 4$ transition probability matrix,
\begin{eqnarray}
\mathbf{P}=\begin{bmatrix}
0.9811 & 0.0189 & 0 & 0\\
0.0250 & 0.9500 & 0.0250 & 0\\
0 & 0.01429 & 0.7143 & 0.1429\\
0 & 0 & 0.0125 & 0.9875\\
\end{bmatrix}, 
\label{eqn:state-transition-matrix}
\end{eqnarray}
is constructed. According to \eqref{eqn:state-transition-matrix}, the highest probabilities occur on the diagonal of the matrix. Hence, if the current solar irradiance is known, it is most likely that the solar intensity of the next time instant would be in a similar range~\cite{Shamshad-Energy:2005}. Now, before applying a Markov model to generate the synthetic data, we need to investigate the state dependency of the chain, i.e., whether or not the successive solar states are dependent upon each other.
\subsection{State dependency test}
\begin{theorem}
For the given time series data set, the successive states are dependent upon each other, and thus a Markov chain can be constructed from it.
\label{theorem:1}
\end{theorem}
\begin{proof}
First, we assume that the null hypothesis holds for the given observed data set. Therefore, the statistics $\alpha$, where
\begin{eqnarray}
\alpha = 2\sum_{i,j}^k n_{i,j}\ln\left(\frac{p_{i,j}}{p_j}\right),
\label{eqn:alpha}
\end{eqnarray}
of the measured data set can be assumed to be distributed asymptotically as $\chi^2$ having $(k-1)^2$ degrees of freedom~\cite{Shamshad-Energy:2005}. In \eqref{eqn:alpha},  $k$ is the total number of states, $n_{i,j}$ is the frequency in state $i$ followed by state $j$, and $p_j$ is the marginal probability for the $j$th column of $\mathbf{P}$ defined as
\begin{eqnarray}
p_j = \frac{\sum_{i}^k n_{i,j}}{\sum_{i,j}^k n_{i,j}}.
\end{eqnarray}
The value of $\alpha$ is determined considering the average of whole time-series data set in summer for the duration $4.30$ am to $7.30$ pm. The value is found to be $374.66$, which is significantly higher than $\chi^2$ value of $16.9189$ at $5\%$ level with $9$ degrees of freedom\footnote{The value of $\chi^2$ at level $5\%$ with $(4-1)^2,\text{i.e.,}~ 9$ degrees of freedom, is calculated via the javascript program available online at http://bavard.fourmilab.ch/rpkp/experiments/analysis/chiCalc.html.}. Hence, the null hypothesis that the successive transitions are independent is rejected. Therefore, the state transition of the observed data depends on previous states, and thus possesses the property of a first order Markov chain. Thus, Theorem~\ref{theorem:1} is proved.
\end{proof}
However, in general, solar irradiance changes significantly with time. For instance, solar irradiance is higher during noon compared early morning and late afternoon. In this regard, we now investigate whether state transition probabilities are affected by the change of time across a day.
\subsection{Temporal stationarity test}
\begin{property}
The Markov chain constructed from the observed data is heterogeneous over time, and thus the probability of state transition changes from one time segment to the next.
\label{theorem:2}
\end{property}
To proof this property, first we divide the whole time series data set into five different intervals where each interval comprises three hours. Now, the Markov chain will possess the temporal stationarity property if the transition probability matrices for the solar data of each time interval are approximately equal to each other~\cite{Shamshad-Energy:2005}. To that end, the computed five transition probability matrices\footnote{We assume $p_{i,j} = \frac{n_{i,j}}{\sum_j n_{i,j}}=0, i = 1, 2,\hdots, 4;~ \forall n_{i,j}=0$ in designing all transition probability matrices.} $\mathbf{P}_1, \mathbf{P}_2, \mathbf{P}_3, \mathbf{P}_4$ and $\mathbf{P}_5$ for the considered five segments of time are shown in Table~\ref{table:2}.
\begin{table*}[t]
\scriptsize
\centering
\renewcommand{\arraystretch}{1.3}
\caption{Table of different transition probability matrices for different intervals of the time series solar irradiance data.}
\begin{tabular}{c c c c c}
\hline
$\mathbf{P}_1=\left[\begin{array}{ c c c c}0.9615 & 0.0385 & 0 & 0\\0 & 1.0 & 0 & 0\\0 & 0 & 0 & 0\\0 & 0 & 0 & 0\end{array}\right]$ & $\mathbf{P}_2=\left[\begin{array}{ c c c c}1 & 0 & 0 & 0\\0 & 0.75 & 0.25 & 0\\0 & 0 & 0.80 & 0.20\\0 & 0 & 0 & 1\end{array}\right]$ & $\mathbf{P}_3=\left[\begin{array}{ c c c c}1 & 0 & 0 & 0\\0 & 1 & 0 & 0\\0 & 0 & 1 & 0\\0 & 0 & 0 & 1\end{array}\right]$ \\ $\mathbf{P}_4=\left[\begin{array}{ c c c c}1& 0& 0 & 0\\0 & 1 & 0 & 0\\0 & 0.5 & 0.5 & 0\\0 & 0 & 0.0588 & 0.9412\end{array}\right]$ & $\mathbf{P}_5=\left[\begin{array}{ c c c c}1.0& 0 & 0 & 0\\0.1111 & 0.8889 & 0 & 0\\0 & 0 & 1 & 0\\0 & 0 & 0 & 1\end{array}\right]$\\
\hline
\end{tabular}
\label{table:2}
\end{table*}
As can be seen from the table, all transition matrices are different, and considerably unequal, to each other. Hence, the probability of transition of states of the Markov chain changes over time.

\subsection{Multiple segment Markov Chain}
In this section, we propose a multiple segment Markov chain approach to capture the variability of the transition probability matrix between different times. We propose that the total time duration of interest will be divided into  multiple time segments $S=|{\mathcal{S}}|$ with the same number of states, where the duration $T_s$ of each segment $s\in\mathcal{S}$ does not need to be equal to other segments. Secondly, the probability transition matrix $\mathbf{P}_s$ is calculated for each $s$ from the time series data. It is assumed that the initial state of the chain is known for the first time segment. For the rest of the segments, the final state of the previous segment will be considered as the initial state of the next.

Now, for a given state transition matrix $\mathbf{P}_s$ of time segment $s\in\mathcal{S}$, the distribution over states at different time instants $t$ of the segment can be expressed as a stochastic row vector $\mathbf{x_s}$ with the relation~\cite{Mbook:1993}
\begin{eqnarray}
\mathbf{x}_s^{t+1} = \mathbf{x}_s^t\mathbf{P}_s.
\label{eqn:transition-of-state}
\end{eqnarray}
For instance, if at time $t$ of segment $s$ the solar irradiance state is $x_\text{M}$ (from \eqref{eqn:states}) the probability of the irradiance state at time $t+3$ can be defined as
\begin{eqnarray}
\mathbf{x}_s^{t+3} &=& \mathbf{x}_s^{t+2}\mathbf{P}_s\nonumber\\&=&\mathbf{x}_s^{t+1}\mathbf{P}^2\nonumber\\&=&\mathbf{x}_s^t\mathbf{P}_s^3\nonumber\\&=&\left[0~ 1~ 0~ 0\right]\mathbf{P}_s^3.
\label{eqn:state-at-t}
\end{eqnarray}
After determining the probability distribution over states at any time $t$, the actual state is assumed to be the state with highest probability at that time. That is
\begin{eqnarray}
x_s^t = \arg(\max(\mathbf{x}_s^t)).
\end{eqnarray}
The steps of determining solar irradiance states using a multiple segment Markov chain process are detailed in Algorithm~\ref{algorithm:1}.
\begin{algorithm}[t]
\caption{Algorithm to determine solar irradiance state via a multiple segment Markov chain.}
\label{algorithm:1}
\begin{algorithmic}[1]
\STATE \textbf{Initialization:} Number of segments $\left[1,2,\hdots,S\right]$. %
\STATE \textbf{Initialization:} Time slots in each segment $\left[T_1,T_2,\hdots,T_S\right]$.%
\STATE \textbf{Initialization:} Vector of transition probability matrices $\left[\mathbf{P}_1, \mathbf{P}_2,\hdots,\mathbf{P}_s\right]$.
\STATE \textbf{Initialization:} Vector of states $\left[x_\text{L},~x_\text{M},~x_\text{H},~x_\text{VH}\right]$.%
\STATE \textbf{Initialization:} Initial state vector $\mathbf{x}_s^\text{ini}$.%
\FOR{Each state $s\in\mathcal{S}$}
	\IF {$s = 1$}
		\STATE $\mathbf{x}_s^{t=1} = \mathbf{x}_s^\text{ini}$
	\ELSE
		\STATE $\mathbf{x}_s^{t=1} = \mathbf{x}_{s-1}^{T_{s-1}}$
	\ENDIF
	\FOR{Each time $t = 2, 3,\hdots, T_s$}
		\STATE $\mathbf{x}_s^{t} = \mathbf{x}_s^{t-1}\times(\mathbf{P}_s)^t$
		\STATE $x_s^t = \arg\max(\mathbf{x}_s^t)$
	\ENDFOR
\ENDFOR
\end{algorithmic}
\end{algorithm}
\section{Case Study}
\begin{figure}[t!]
\centering
\includegraphics[width=\columnwidth]{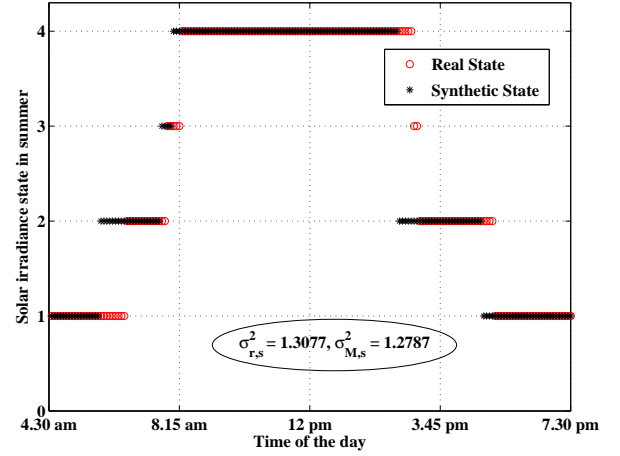}
\caption{Comparison of real solar states and synthetic solar states generated by the proposed multiple segment Markov chain model in summer.} \label{fig:solar-states-summer-prop}
\end{figure}

\begin{figure}[t!]
\centering
\includegraphics[width=\columnwidth]{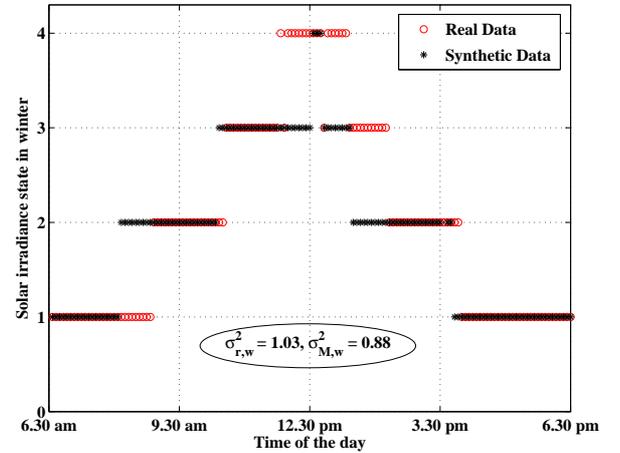}
\caption{Comparison of real solar states and synthetic solar states generated by the proposed multiple segment Markov chain model in winter.}
\label{fig:solar-states-winter-prop}
\end{figure}

To show the effectiveness of the proposed scheme to generate synthetic states of solar irradiance at different times of the day, we conduct numerical experiments for two cases: 1) for average time series solar data in summer and 2) for average time series solar data in winter. Both sets of data are collected from South Park, Colorado, and are accessible online at \cite{NREL-2014}. Based on the time series data, a total of four irradiance states are considered as in \eqref{eqn:states}. In Table~\ref{table:3}, we show different threshold values chosen to define different states.
\begin{table}[h]
\centering
\caption{Threshold values for defining different solar irradiance state according to \eqref{eqn:states} and Table \ref{table:1}.}
\begin{tabular}{|c|c|}
\hline
Thresholds & $I_s$ in W/m$^2$\\
\hline\hline
$I_{s,\text{lr}}$ & $200$ \\
\hline
$I_{s,\text{mr}}$ & $450$ \\
\hline
$I_{s,\text{max}}$& $500$\\
\hline
\end{tabular}
\label{table:3}
\end{table}
We choose the time duration $4.30$ am to $7.30$ pm for summer, and $6.00$ am to $6.00$ pm for winter as the total period of time to generate the synthetic data using the proposed scheme. We note that the solar irradiance data beyond these two intervals are too small to produce any meaningful state transition matrix. We divide the total time interval into five different segments for summer and four segments for winter according to Table~\ref{table:4}.
\begin{table}[h]
\centering
\caption{Duration of different time segments in different seasons.}
\small
\begin{tabular}{|c|c|c|c|c|}
\hline
Time segment & Summer & Winter\\
\hline\hline
$s_1$ & 4.30 am to 7.30 am & 6.00 am to 9.00 am\\
\hline
$s_2$ & 7.30 am to 10.30 am & 9.00 am to 12.00 pm \\
\hline
$s_3$ & 10.30 am to 1.30 pm & 12.00 pm to 3.00 pm \\
\hline
$s_4$ & 1.30 pm to 4.30 pm & 3.00 pm to 6.00 pm\\
\hline
$s_5$ & 4.30 pm to 7.30 pm & - \\
\hline
\end{tabular}
\label{table:4}
\end{table}
The state transition matrices for different segments of time are computed, and solar irradiance states are determined using Algorithm~\ref{algorithm:1}. The experiments are conducted for the proposed scheme to generate the synthetic states, and are compared to the real measured states in Fig.~\ref{fig:solar-states-summer-prop} and Fig.~\ref{fig:solar-states-winter-prop}.

In Fig.~\ref{fig:solar-states-summer-prop}, we show the relation between the real observed solar states with the generated synthetic solar states via the proposed multiple segment Markov approach, in summer. It can be seen that the solar states gradually increases from state one to state four as the time of the day increases from $4.30$ am to $1.30$ pm, and again moves back to state one eventually around $6.00$ pm, which is in fact due to the sunrise and sunset times of the location.  However, according to Fig.~\ref{fig:solar-states-summer-prop}, the synthetic states generated via the proposed scheme have noticeably good performance in resembling real states, apart from minor deviations during early morning and early afternoon. According to the data set for summer, the standard deviation $\sigma_{M,s}^2$ of the generated synthetic states is $1.2787$, which is almost equal to the standard deviation of the real observed set of states $\sigma_{r,s}^2 = 1.3077$. Hence, according to the performance in Fig.~\ref{fig:solar-states-summer-prop}, the proposed multiple segment Markov chain can be successfully used to produce synthetic solar data for summer-time.

To demonstrate how effectively the proposed scheme can capture solar variation in winter, we show the same comparison for the time series data set of winter in Fig.~\ref{fig:solar-states-winter-prop}. We note that in calculating the states in winter we reduce the number of segments of total duration of time into four as the changes in solar irradiance are not significant enough to build transition probability matrices for more than four segments. As can be seen from Fig.~\ref{fig:solar-states-winter-prop}, the performance of the proposed multiple segment Markov chain is affected by this change in solar irradiance. Hence, importantly, the generated states have relatively poorer performance than in summer. In fact, in winter, the solar conditions change more abruptly than in summer, including late sunrise and early sunset. Hence, to generate states that match the real states well is very challenging. However, except for a few time steps around noon, the proposed generator produces solar states for winter which are closely affine to the real states as can be seen from Fig. \ref{fig:solar-states-winter-prop}.  According to Fig.~\ref{fig:solar-states-winter-prop}, for winter, the standard deviation $\sigma_{r,w}^2$ of real data set is $1.03$, which is close to the standard deviation $\sigma_{M,w}^2 = 0.88$ of the generated synthetic data. Therefore, considering the performance based on the considered data set, it can be concluded that the proposed multiple segment Markov chain approach is applicable to generate solar states in both summer and winter seasons.

\section{Conclusion}\label{sec:conclusion}
In this paper, a multiple segment Markov chain is studied to produce synthetic solar states at different times of the day for both summer and winter seasons. With a real time series data set, it has been shown theoretically that solar states are interdependent, and thus can be modeled by a Markov approach. It has further been proved that the probability of transition of one solar state to another state depends on time, and therefore the total time of interest is divided into multiple segments to capture this temporal dependency. Based on real solar data sets from Colorado, a probability transition matrix has been calculated for each of the segments, and an algorithm has been proposed to determine the state for each time instant of the day. The effectiveness of the proposed scheme has been demonstrated via numerical simulation, with noticeably good performance in terms of the resemblance of generated synthetic solar states with the real data set.

The proposed scheme can be extended and improved in various aspects. The number of states in the model can be better calibrated using real data from different locations, and thus can be used to propose a generic state derivation model. The theoretical relationship between the number of time segments and solar irradiance at different time slots is also worthy of further investigation. Further, by introducing a learning capability, the proposed scheme can be extended to efficiently characterize the inter-temporal behavior of the solar irradiance. %
\def\baselinestretch{0.82}

\end{document}